\documentclass[11pt,a4paper]{article}
\usepackage[latin1]{inputenc}
\usepackage{amsmath}
\usepackage[english]{babel}
\usepackage{amsfonts}
\usepackage{amssymb}
\usepackage{graphicx}
\usepackage{hyperref}

\author{Valerio Lucarini$^*$$^\dagger$ \\
\texttt{valerio.lucarini@zmaw.de},\\
Tobias Kuna$^\dagger$ \\
\texttt{t.kuna@reading.ac.uk},\\
Jeroen Wouters$^*$ \\
\texttt{jeroen.wouters@zmaw.de},\\
Davide Faranda$^*$ \\
\texttt{davide.faranda@zmaw.de}\\
}
\title{Relevance of sampling schemes in light of Ruelle's linear response theory}

\begin{document}

\maketitle

\begin{center}
\begin{footnotesize}
\textit{$^*$Klimacampus, Universit\"at Hamburg, Hamburg, Germany\\ $^\dagger$Department of Mathematics and Statistics, University of Reading, Reading, UK}
\end{footnotesize}
\end{center}

\abstract{We reconsider the theory of the linear response of non-equilibrium steady states to perturbations. We first show that by using a general functional decomposition for space-time dependent forcings, we can define elementary susceptibilities that allow to construct the response of the system to general perturbations. Starting from the definition of SRB measure, we then study the consequence of taking different sampling schemes for analysing the response of the system. We show that only
a specific choice of the time horizon for evaluating the response of the system to a general time-dependent perturbation allows to
obtain the formula first presented by Ruelle. We also discuss the special case of periodic perturbations, showing that when they are
taken into consideration the sampling can be fine-tuned to make the definition of the correct time horizon immaterial. Finally, we discuss the implications of our
results in terms of strategies for analyzing the outputs of
numerical experiments by providing a critical review of a formula proposed by Reick.}

\section{Introduction}

The study of how the properties of general
non-equi\-librium statistical mechanical systems change when
considering a ge\-ne\-ric perturbation, usually related to variations
either  in the value of some internal parameters or in the
external forcing, is of great relevance, both in purely
mathematical terms and with regards to applications to the natural
and social sciences. Whereas in quasi-equilibrium statistical
mechanics it is possible to link the response of a system to
perturbations to its unforced fluctuations thanks to the
fluctuation-dissipation theorem \cite{kubo_66,zubarev}, in the
general non-equilibrium case it is not possible to frame
rigorously an equivalence between internal fluctuations and
forcings. At a fundamental level, this is closely related to the
fact that forced and dissipative systems feature a singular
invariant measure. Whereas natural fluctuations of the system are
restricted to the unstable manifold, because, by definition,
asymptotically there is no dynamics along the stable manifold,
perturbations will induce motions - of exponentially decaying
amplitude - out of the attractor with probability one, as
discussed in, e.g., \cite{ruelle_review_2009,ruelle_98,luc08}. It
is worth noting that Lorenz anticipated some of these ideas when
studying the difference between free and forced variability of the
climate system \cite{lorenz79}. This crucial difficulty inherent
to out-of-equilibrium systems is lifted if the external
perturbation is, rather artificially, everywhere tangent to the
unstable manifold, or if the system includes some stochastic
forcing, which smooths out the resulting the invariant measure
\cite{laco07}.

Recently, Ruelle
\cite{ruelle_differentiation_1997,ruelle_98,ruelle_review_2009}
paved the way to the study of the response of general
non-equilibrium systems to perturbations by presenting rigorous
results leading to the formulation of a response theory for Axiom
A dynamical systems \cite{ruelle89}, which possess
a Sinai-Ruelle-Bowen invariant measure \cite{young_what_2002}. Given
a measurable observable of the system, the change in its
expectation value due to an $\epsilon$-perturbation in the flow
(or in the map, in the case of discrete dynamics) can be written
as a perturbative series of terms proportional to $\epsilon^n$,
where each term of the series can be written as the expectation
value of some well-defined observable over the unperturbed state.
Ruelle's formula is identical to Kubo's classical formula
\cite{kubo_57} when a Hamiltonian system is considered
\cite{luc08}.

Whereas Axiom A systems are mathematically non-generic, the
applicability of the Ruelle theory to a variety of actual models
is supported by the so-called \textit{chaotic hypothesis}
\cite{galla}, which states that systems with many degrees of
freedom behave as if they were Axiom A systems when macroscopic
statistical properties are considered. The chaotic hypothesis has
been interpreted as the natural extension of the classic ergodic
hypothesis to non-Hamiltonian systems \cite{galla06}.

In the last decade great efforts have been directed at extending
and clarifying the degree of applicability of the response theory
for non-equilibrium systems along five main lines:
\begin{itemize}
    \item extension of the theory for more general classes of dynamical
systems \cite{dolgopyat,baladi};
    \item introduction of effective algorithms for computing the response
    in dynamical systems with many degrees of freedom \cite{abra07}, in order
    to support the numerical analyses pioneered by
    \cite{reick_linear_2002,cessac};
    \item investigation of the frequency-dependent response - the
    susceptibility -
    for the linear and nonlinear cases, with the ensuing introduction
    of a new theory of Kramers-Kronig relations and sum rules for non-equilibrium systems
    \cite{luc08,shimizu10} supported by numerical experiments \cite{luc09};
    \item study of the response to external perturbations of
    non-equi\-lib\-rium systems undergoing stochastic dynamics
    \cite{majda_stochastic_2010,yuge10};
    \item use of Ruelle's response theory to study the impact of
    adding stochastic forcing to otherwise deterministic systems
    \cite{luc11b}.
\end{itemize}
In particular, the response theory seems especially promising for
tackling notoriously complex problems such as those related to
studying the response of geophysical systems to perturbations,
which include the investigation of climate
change; see discussions in \cite{abra07,luc09,luc11}. In
particular, in \cite{luc11}, it is discussed that by deriving from
the linear susceptibility the time-dependent Green function, it is
possible to devise a strategy to compute climate change for a
general observable and for a general time-dependent pattern of
forcing. Recently, response theory is becoming of great interest
also in social sciences such as economics \cite{hawkins_2009}.

When developing a response theory, there are two
possible ways to frame the temporal impact of the additional
perturbation to the dynamics. Either one considers the impact
at a given time $t$ of a perturbation affecting the system since a very
distant past, or one considers the impact in the distant future of
perturbations starting at the present time. When deriving the response formula, {Ruelle} takes the first approach and delivers the \textit{correct} formula~\cite{ruelle_differentiation_1997}. Taking a
different point of view and considering the specific case of
periodic perturbations -- which, anyway, tell us the whole story
about the response by linearity --, Reick \cite{reick_linear_2002} derives a formula that is well suited for analyzing the output of numerical experiments~\cite{luc09,luc11}.

Given the great relevance and increasing popularity in
applications of the response theory introduced by Ruelle, in this
paper, we reconsider the theory of the linear response of
non-equilibrium steady states to perturbations and try to bridge the theoretical derivations and the strategies for
designing numerical experiments and analyzing efficiently their
outputs.

In Section 2, we study the relevance of the choice of
the time horizon for evaluating the impact of the perturbation and
we demonstrate by direct calculation that the Ruelle approach is the
correct one. We clarify some of the assumptions implicitly
considered in his derivations. We then discuss the special
case of periodic perturbations, showing that using them as basis
for a response theory greatly simplifies the formulas and the
conditions under which the formulas are derived. In Section 3, we
discuss the implications of our results in terms of strategies for
improving the quality of numerical simulations and of the analysis
of their output signals and reconsider Reick's formula~\cite{reick_linear_2002}. In Section 4 we present our conclusions
and perspectives for future work.

\section{Linear Response Theory, revised}
\label{sec:linear_response_theory}
\subsection{Separable perturbations}
We study the linear response of a discrete dynamical system to
general time-dependent perturbations. All calculations are formal, in the sense that we neglect all higher orders in the perturbation without deriving an estimate for these terms and we assume that all sums converge in all senses necessary.

The unperturbed dynamical system is
given by
\[ x_{t+1} = f(x_t) \,,\]
with $t \in \mathbb{Z}$, $x_t \in M$, $M$ being a smooth manifold
and $f: M \rightarrow M$ a differentiable map. For simplicity we
consider a time-independent unperturbed dynamics, although the
following can be extended to a time-dependent case in a
straightforward manner. Moreover, the analysis of the case of a
continuous time flow $\dot{x} = f(x)$ is perfectly analogous to
what is presented in the following and the main corresponding results will
be mentioned in Appendix \ref{app:continuous}.

The dynamical system is perturbed by a time-dependent forcing
$X(t,x)$ as follows:
\begin{align}
\tilde{x}_{t+1} = \tilde{f}_{t+1}(\tilde{x}_t) := f(\tilde{x}_t) + X(t+1,f(\tilde{x}_t)) \,. \label{eq:perturbation}
\end{align}

The effect of the perturbation on individual trajectories is in
general difficult to describe. More can however be said about the statistical properties of the system. One can look at the expectation
values of observables under invariant states of the dynamical
system:
\begin{align*}
\rho(A) := \int \rho(dx) A(x) \,,
\end{align*}
where $\rho(dx)$ is an invariant measure of the unperturbed dynamics i.e.
\begin{align*}
\rho(A \circ f) = \rho (A) \,.
\end{align*}
for any observable A.
In general, a dynamical system can possess many invariant
measures.  The physically relevant measure
for dynamical systems is the SRB measure~\cite{young_what_2002}. This measure is
physical in the sense that for a set of initial conditions of full
Lebesgue measure the time averages $\lim_{t \rightarrow \infty}
\frac{1}{t} \sum_{k=1}^t A(f^k(x))$ converge to the expectation
value under $\rho$. In other words, for any measure $l(dx)$ that
is absolutely continuous w.r.t. Lebesgue, we have that
\begin{align}
\rho(A) = \lim_{t \rightarrow \infty} \frac{1}{t} \sum_{k=1}^t \int l(dx) A( f^k (x)) \,. \label{eq:SRB}
\end{align}

We want to determine the linear response of expectation values
under the SRB measure to perturbations of the dynamical system
as in Eq.~\ref{eq:perturbation}. We denote by $\delta_T \rho$ the difference in the expectation
value between the perturbed and unperturbed system at time $T$.
In~\cite{ruelle_review_2009}, Ruelle presents a formula for the
linear response due to perturbations that are separable in time and space:
\begin{align*}
X(t,x) = \phi(t) \chi(x) \,.
\end{align*}
The leading order term of the expansion of $\delta_T \rho(A)$ in $X$ is given by
\begin{align}
\delta_T \rho(A) \approx \sum_{j\in \mathbb{Z}} G_A(j) \phi(T-j) \,, \label{eq:ruelle_time_sep}
\end{align}
with
\begin{align}
G_A(j)= \theta(j) \int \rho(dx) \chi(x) D(A \circ f^j)(x) \,, \label{eq:response_func}
\end{align}
where $\theta$ is the Heaviside function. Since $\delta_T \rho(A)$ is expressed as a convolution product of $G_A$ and $\phi$, the Fourier transform of the response $\delta_\omega \rho (A) =
\sum_{T \in \mathbb{Z}} e^{i T \omega} \delta_T \rho (A)$ is given
by a product of the Fourier transform $\hat{\phi}(\omega)$ of the time factor $\phi(t)$
and a susceptibility function $\hat{\kappa}_A (\omega)$:
\begin{align}
\delta_\omega \rho (A) &\approx \hat{\kappa}_A (\omega) \hat{\phi} (\omega) \,. \label{eq:response}
\end{align}
where
\begin{align}
\hat{\phi}(\omega)&=\sum_{j \in \mathbb{Z}} e^{i j \omega} \phi(j) \,, \nonumber \\
\hat{\kappa}_A(\omega) &=  \sum_{j \in \mathbb{Z}} e^{i j \omega}  G_A(j) \nonumber \\
&= \sum_{j \geq 0} e^{i j \omega} \int \rho (dx) \chi(x) D(A \circ f^j)(x) \,.
\label{eq:suscep}
\end{align}
Due to the causality of the response function $G_A(j)$
(i.e. $G_A(j)=0$, $j< 0)$), the susceptibility $\hat{\kappa}_A(\omega)
$ is analytic in the upper complex plane and satisfies
Kramers-Kronig relations~\cite{ruelle_98,luc08}.

\subsection{General perturbations}
If the perturbation is of a more general nature (i.e. not
separable), we can deduce a linear response formula from
Eq.~\ref{eq:response}, solely based on linearity in the following way. Let $\phi_r(t)$ be a Schauder basis~\cite{lindenstrauss_classical_1996} of
time-dependent functions and $\psi_s(x)$ a Schauder basis of
space-dependent functions. One can take for example the Fourier basis in time and a wavelet basis in space, or whatever basis may be suitable for the system at hand. The product functions $\phi_r(t)
\psi_s(x)$ then form a basis of the time and
space dependent functions as a tensor product~\cite{ryan_2002}. More concretely, we may for an appropriate sense of convergence assume that any function $X(t,x)$ can be decomposed in the product basis $\phi_r(t)
\psi_s(x)$ with coefficients $a_{r,s}$:
\begin{align*}
X(t,x) = \sum_{r,s \geq 0} a_{r,s} \phi_r(t) \psi_s(x) \,.
\end{align*}
Since each of the factors in this sum is separable, we can  use
Eq.~\ref{eq:response} and the linearity of the response to get
that the response is given by
\begin{align}
 \delta_\omega \rho (A) \approx \sum_{r,s \geq 0} a_{r,s}
\hat{\phi}_r(\omega) \hat{\kappa}_{s,A}(\omega) \,, \label{eq:suscav}
\end{align}
where $\hat{\kappa}_{s,a}$ is the susceptibility function of observable $A$,
corresponding  to the forcing pattern given by $\psi_s(x)$. Since the vectors $\psi_s(x)$ constitute a basis, the functions $\hat{\kappa}_{s,a}$ are \emph{elementary} linear susceptibilities that allow to construct the response of the system to any pattern of forcing.

By inserting the expression of $\hat{\kappa}_{s,A}(\omega)$ from Eq. \ref{eq:suscep} into Eq.
\ref{eq:suscav}, it is possible to deduce the frequency-dependent response of the system. It is expressed as an ensemble average of a dot product of Fourier transforms, namely the transforms of the perturbation term and of the linear tangent of the observable, $G_A(\omega,x)$:
\begin{align}
\delta_\omega \rho (A) &\approx \sum_{j \geq 0} e^{i j \omega} \int \rho (dx) \hat{X}(\omega,x)
D(A \circ f^j)(x) \nonumber
\\
&= \int \rho(dx) \hat{X}(\omega,x) G_{A}(\omega,x) \,, \label{eq:suscav2}
\end{align}
with
\begin{align}
G_{A}(\omega,x) &= \sum_{j \geq 0} e^{i j \omega} D(A \circ f^j)(x) \nonumber \\
\hat{X}(\omega,x) &= \sum_{T \in \mathbb{Z}} X(T,x) e^{i \omega T} \nonumber \\
&= \sum_{r,s \geq 0} a_{r,s} \hat{\phi}_r (\omega) \psi_s (x). \label{eq:suscav3}
\end{align}
Instead from Eqs.~\ref{eq:ruelle_time_sep}-\ref{eq:response_func} in the time domain
\begin{align}
\delta_T \rho(A) \approx \int \rho(dx) \sum_{j\geq 0} X(T-j,x) D(A \circ f^j)(x) \,. \label{eq:ruelle_time}
\end{align}
For the case of a periodic perturbation $X(t+\tau,x)=X(t,x)$, where $\tau\in\mathbb{N}$, we get as linear response
\begin{align}
\delta_T \rho (A) &\approx \int \rho(dx) \sum_{n=1}^\tau \sum_{m=0}^\infty X(T-n-m \tau,x) D(A \circ f^{n+m \tau})(x) \nonumber \\
%&= \int \rho(dx) \sum_{n=1}^\tau X(T-n,x) \sum_{m=0}^ \infty D(A \circ f^ {n + m \tau}) (x) \nonumber \\
&= \int \rho(dx) \sum_{n=1}^\tau X(T-n,x) G_{A,n} (x) \,, \label{eq:ruelle_periodic}
\end{align}
with
\begin{align*}
G_{A,n} (x) = \sum_{m=0}^ \infty D(A \circ f^ {n + m \tau}) (x) \,.
\end{align*}

In order to elucidate some crucial aspects of the Ruelle's
response theory, we now propose a direct derivation of the linear
response to the perturbation $X(t,x)$ by considering the history
of the perturbed and unperturbed trajectory of the system and
verify under which conditions we find agreement with Eqs.
\ref{eq:suscav2}-\ref{eq:ruelle_time}. Our goal is to derive the leading order term of the
expansion of $\delta_T \rho(A)$ with respect to $X$ from first principle, i.e. without resorting to the Schauder decomposition as above. Such a derivation should of course arrive at the same results as those in Eqs. \ref{eq:suscav2}-\ref{eq:ruelle_time}.

\subsubsection{Response at a moving time horizon}
\label{sec:forward}
We describe the perturbed measure $\tilde{\rho}_T(A)$ such that the system is initialized at time $T$ in an initial condition according to the measure $l$. We move the time horizon at which we observe forward and average the time-evolved measurements. The system is prepared and then observed while it is evolving over a sufficiently long time. The measure $\tilde{\rho}_T$ is time-dependent as the dynamics $\tilde{f}$
is also time-dependent.
Formally we take $\tilde{\rho}_T$ to be the ergodic mean of the expectation values of $A$, starting at time $T$:
\begin{align}
\tilde{\rho}_T (A) &= \lim_{t \rightarrow \infty}
\frac{1}{t}\sum_{k=1}^t \int l(dx) A(\tilde{f}_T^{k}(x)) \,. \label{eq:time_SRB}
% \tilde{\rho}_T(dx) &= \lim_{t \rightarrow \infty}
% \frac{1}{t}\sum_{k=1}^t (\tilde{f}_T^{k})^* l(dx)
\end{align}
Here $l(dx)$ is an initial measure that is absolutely continuous
with respect to Lebesgue and $\tilde{f}^k_T$ represents $k$
iterations of the perturbed dynamics from time $T$ to $T+k$:
\begin{align}
\tilde{f}^k_T (x) = \tilde{f}_{T+k} \circ \ldots \circ \tilde{f}_{T+1} (x) \label{eq:forward} \,.
\end{align}
The difference in expectation values $\delta_T \rho$ is the given by
\begin{align}
 \delta_T \rho(A) = \tilde{\rho}_T(A) - \rho(A). \label{eq:deltarho}
\end{align}

Following the computation presented in \cite{ruelle_review_2009} for the separable case, we can expand the perturbed dynamics $\tilde{f}$ around the
unperturbed dynamics $f$. We then try to rewrite the response of the perturbed system in terms of the SRB measure of the unperturbed system by finding an expression for $A(\tilde{f}_T^k(x))$ in terms of $A(f^k(x))$.

We can approximate up to
first order in $X$ the two time step future evolution by expanding around the
unperturbed dynamics $f^2(x)$:
\begin{align*}
\tilde{x}_{T+2} &= \tilde{f}_{T+2} \circ \tilde{f}_{T+1} (\tilde{x}_T) \\
 & \approx f^2(\tilde{x}_T) + X(T+1,f(\tilde{x}_T)).Df(f(\tilde{x}_T)) + X(T+2,f^2(\tilde{x}_T)) \,.
\end{align*}
For $k$ time steps we similarly get:
\begin{align*}
\tilde{x}_{T+k} &= \tilde{f}_{T+k} \circ \ldots \circ \tilde{f}_{T+1} (\tilde{x}_T) \\
&\approx f^k (\tilde{x}_T) + \sum_{j=1}^k X(T+j,f^{j}(\tilde{x}_T)).(Df^{k-j})(f^{j} (\tilde{x}_T) )\,.
\end{align*}
Thus, we can approximate $A(\tilde{f}_{T+k} \circ \ldots \circ \tilde{f}_{T+1} (x))$ to first order in $X$ as follows:
\begin{align}
A(\tilde{f}_{T+k} \circ \ldots \circ \tilde{f}_{T+1} (x)) \approx & A(f^k(x)) \nonumber \\
&+ A^\prime (f^k(x)) \left( \sum_{j=1}^k X(T+j,f^{j}(x)).(Df^{k-j})(f^{j} (x) ) \right) \nonumber \\
&= A(f^k(x)) \nonumber \\
&+ \sum_{j=1}^{k} X(T+j,f^{j}(x)) D(A \circ f^{k-j}) (f^{j}(x)) \,. \label{eq:A_approx}
\end{align}

The linear response of $A$ is obtained by substituting  Eq.
\ref{eq:A_approx} into Eq. \ref{eq:deltarho}, through Eq.~\ref{eq:SRB} and Eq.~\ref{eq:time_SRB}:
\begin{align*}
\delta_T \rho(A) &\approx \lim_{t \rightarrow \infty} \frac{1}{t} \sum_{k=1}^t \int l(dx) \sum_{j=1}^{k} X(T+j,f^{j}(x)) D(A \circ f^{k-j}) (f^{j}(x)) \nonumber  \\
%& = \lim_{t \rightarrow \infty} \frac{1}{t} \sum_{j = 1}^{t} \sum_{i=0}^{t-j}  \int l(dx) X(T+j,f^j(x)) D(A \circ f^{i}) (f^j(x)) \nonumber \\
& = \lim_{t \rightarrow \infty} \frac{1}{t} \sum_{i = 0}^{t-1} \sum_{j=1}^{t-i}  \int l(dx) X(T+j,f^j(x)) D(A \circ f^{i}) (f^j(x)) \,. \nonumber
\end{align*}
Using that for $ i \geq t $ the expression is zero, we have
\begin{align}
\delta_T \rho(A) & \approx \sum_{i \geq 0} \int \left( \lim_{t \rightarrow \infty} \frac{1}{t}
\sum_{j=1}^{t-i} (f^*)^{j} l(dx) X(T+j, x) \right) D(A \circ f^{i}) (x) \,. \label{eq:future}
\end{align}
Note that it is not possible to rewrite the sum in
$j$ as the ergodic time mean of $l$ due to the time dependence of the
perturbation $X(T+j,x)$. Therefore, surprisingly, Eq. \ref{eq:future} does not in general
agree with Eq.~\ref{eq:ruelle_time}.
In particular, by taking the
limit on the right hand side, we obtain that the
$T$-dependence disappears. Say we shift $T$ to $T-T^\prime$ in the limit appearing in the above equation:
\begin{align*}
&\lim_{t \rightarrow \infty} \frac{1}{t} \sum_{j=1}^{t-i} (f^*)^{j} l(dx) X(T-T^\prime + j, x) \\
&= \lim_{t \rightarrow \infty} \frac{1}{t} \sum_{j^\prime = 1-T^\prime}^{t-i-T^\prime} (f^*)^{j^\prime + T^\prime} l(dx) X(T + j^\prime, x) \\
&= \lim_{t \rightarrow \infty} \frac{1}{t} \sum_{j^\prime = 1}^{t-i} (f^*)^{j^\prime} \left((f^*)^{T^\prime} l(dx) \right) X(T + j^\prime, x) \,.
\end{align*}
Taking the reasonable assumption that the result in Eq.~\ref{eq:future} does not depend on the initial measure $l(dx)$ (this cannot be obtained from the uniqueness of the SRB measure), the obtained response of the
system is time-independent even if the
forcing is time-dependent.

Let us compare the result contained in Eq.~\ref{eq:future} with Eq.~\ref{eq:ruelle_time_sep} in the special case of a
time-independent perturbation $X(t,x)=\chi(x)$. Now Eq.~\ref{eq:future}
and Eq. \ref{eq:ruelle_time_sep} agree since Eq.~\ref{eq:future} simplifies to:
\begin{align*}
\delta_T \rho (A) \approx \sum_{i \geq 0} \int \rho(dx) \chi(x) D(A\circ
f^i)(x) \,,
\end{align*}
because $\lim_{t \rightarrow \infty} 1/t \sum_{j=1}^{t-i}
(f^*)^{j} l(dx))=\rho(dx)$, by the definition of the SRB measure. The formula given by Ruelle
\cite{ruelle_98} is recovered, as can be seen by substituting $\phi(t)=1$ into Eq.~\ref{eq:ruelle_time_sep}.

However, already in the case of a time-periodic perturbation \[X(t,x)=X(t+\tau,x)\] there is no agreement between Eq.~\ref{eq:future} and Eq.~\ref{eq:ruelle_time}. In this case the sum over $j$ appearing in
Eq.~\ref{eq:future} can be written as a double sum, one over $k$ periods, indexed by $m$, and one over the $\tau$ phases in each period, indexed by $n$:
\begin{align}
& \lim_{t \rightarrow \infty} \frac{1}{t} \sum_{j=1}^t (f^*)^j l(dx) X(T+j,x) \nonumber \\
& = \lim_{k \rightarrow \infty} \frac{1}{k \tau} \sum_{m=1}^k \sum_{n=1}^\tau (f^*)^{m \tau} (f^*)^{n} l(dx) X(T+n,x) \nonumber \\
& = \frac{1}{\tau}\sum_{n=1}^\tau \rho_n(dx) X(t+n,x) \,, \label{eq:moving_periodic}
\end{align}
with
\[ \rho_n(dx) = \lim_{k \rightarrow \infty} \frac{1}{k } \sum_{m=1}^k (f^*)^{m \tau} (f^*)^{n} l(dx) \,. \]
Under the assumption that $\rho_n = \rho$ for all $n \in \lbrace 1,\ldots,\tau \rbrace $, i.e. sub-sampling does not impact the unperturbed invariant measure, the response gives a similar result as the Ruelle formula, but with an averaged perturbation. Substituting Eq.~\ref{eq:moving_periodic} into Eq.~\ref{eq:future}, we obtain a formula of the form of Eq.~\ref{eq:ruelle_time}, with the difference that instead of the true forcing $X(t,x)$ the averaged forcing
\[\frac{1}{\tau} \sum_{n=1}^\tau X(t+n,x) \]
appears. The disagreement is apparent, e.g. when one considers a perturbation of the form $X(t,x)=\sin(\frac{2\pi l}{\tau}t) \chi(x)$,
which obviously results in a zero response. This effect has a clear intuitive interpretation. The response at a given time depends mostly on the immediate past, hence if one does not keep fixed the horizon, one risks to average out the variability. The previous formula reflects this intuition. 

One way to obtain agreement with Formula \ref{eq:ruelle_periodic} is to choose a specific sampling procedure. We sample with the same periodicity $\tau$ of the forcing, thus altering the definition of the response. We define the measures for the perturbed and unperturbed system as
\begin{align}
\tilde{\rho}^\prime_{T,p} (dx) &:= \lim_{N \rightarrow \infty} \frac{1}{N} \sum_{k=0}^N (\tilde{f}^{k \tau + p}_T)^* l(dx) \nonumber \\
\rho^\prime (dx) &:= \lim_{N \rightarrow \infty} \frac{1}{N} \sum_{k=0}^N (f^{k \tau + p})^* l(dx) \,. \label{eq:periodic_sampling}
\end{align}
With this definition we obtain using Eq.~\ref{eq:A_approx}:
\begin{align*}
\delta \rho^\prime_{T,p} (A) &= \tilde{\rho}^\prime_{T,p} (A) - \rho^\prime (A) \nonumber\\
& \approx \lim_{N \rightarrow \infty} \frac{1}{N} \sum_{k=0}^N \left( \sum_{m=-1}^{N-1} \sum_{i=1}^{N-m} \theta(m \tau + n + p) \right) \int \left( f^{m \tau + n + p} \right)^* l (dx) \nonumber \\
& \hspace{3cm} X(T+ m \tau + n + p, x ) D(A \circ f^{k \tau - m \tau - n})(x) \,. \nonumber
\end{align*}
Using the periodicity of $X$ one can obtain
\begin{align*}
%&= \sum_{n=1}^\tau \lim_{N \rightarrow \infty} \frac{1}{N} \sum_{m=0}^{N-1} \sum_{k=m+1}^N \int (f^{n+m \tau})^* l(dx) X(T+n, x) D(A \circ f^{(k-m) \tau -n})(x) \nonumber \\
%\delta \rho^\prime_T  &\approx \sum_{n=1}^\tau \lim_{N \rightarrow \infty} \frac{1}{N}  \sum_{i \geq 1} \sum_{m=0}^{N-i} \int (f^{n+m \tau})^* l(dx) X(T+n, x) D(A \circ f^{i \tau -n})(x) \nonumber \\
\delta \rho^\prime_{T,p} (A) &\approx \sum_{n=1}^\tau \sum_{i\geq 1} \int \left( \lim_{N \rightarrow \infty} \frac{1}{N}  \sum_{m=-1}^{N-i} (f^{n+m \tau+p})^* l(dx) \theta(m \tau + p +n) \right) \\
& \qquad \quad X(T+n+p, x) \left( D(A \circ f^{i \tau -n})(x) \right) \nonumber \\
&= \sum_{n=0}^{\tau-1} \int \rho(dx) X(T+p-n,x) G_{A,n} (x) = \delta_{T+p} \rho(A) \,. \nonumber
\end{align*}
Hence by choosing the initial phase $p$ at which we start sampling, we can obtain the response at this phase. This means that we only need to start one long simulation of $f$ and $\tilde{f}$ and do summations of the differences $(A\circ \tilde{f} - A\circ f)(x)$ according to Eq.~\ref{eq:periodic_sampling} at all phases $p$ in one period to obtain the entire response to the periodic forcing. By applying a forcing that contains several frequencies, such as a block wave, we can extract the susceptibility at all present frequencies in one run by taking the Fourier transform of the response. 

Note that if we sample the signal with a periodicity $\eta$ which is prime with respect to the period $\tau$ of the forcing, we will obtain no $p$-dependence (with $p$, in this case, ranging from 0 to $\eta-1$) in the response. For all values of $p$ we will obtain as a result the response to the time-averaged forcing. Therefore, the case of sampling at all time steps discussed above is just the special case given by $\eta=1$, where we are basically considering the case of the Nyquist frequency. Instead, if $\tau$ and $\eta$ are not prime with respect to each other, the sampling procedure will be able to ascertain the $p$-dependence of the response of the system at the periodicity given by the common harmonic terms.

If the periodicity of the forcing is not known, the above discussion tells us that by doing a sampling at larger and larger periods $\eta$ and checking for each of those the phase-dependence of the response, it is possible to deduce the fundamental period of the forcing. If the procedure does not converge, we are facing a quasi-periodic or continuous-spectrum forcing for which this approach fails.

Therefore, this situation is unsatisfactory. Why do we only get the correct result for periodic perturbations and fine-tuning the sampling or by taking constant perturbations?

\subsubsection{Response at a fixed time horizon}
\label{sec:backward}
This paradox can be resolved by defining the
time-dependent SRB measure in Eq.~\ref{eq:time_SRB} using a different method of sampling. We now consider the time evolution
$\tilde{f}^k_T$ in this definition to go from
time $T-k$ in the past up to the fixed time horizon $T$, so instead of Eq.~\ref{eq:forward}, we have:
\begin{align}
\tilde{f}^k_T = \tilde{f}_{T} \circ \ldots \circ \tilde{f}_{T-k} \label{eq:backward} \,.
\end{align}
Note that this approach does not use the reversed time dynamics but rather a different time perspective in which the final time is fixed as the current time and the perturbation starts in the remote past.

The expansion to first order in $X$ around the dynamics of $f$ now becomes:
\begin{align}
\tilde{x}_T &= \tilde{f}_T \circ \ldots \circ \tilde{f}_{T-k+1} (\tilde{x}_{T-k}) \nonumber \\
            &\approx f^k (\tilde{x}_{T-k}) + \sum_{j=0}^{k-1} X(T-j,f^{k-j}(\tilde{x}_{T-k})).(Df^{j})(f^{k-j} (x_{T-k})) \,. \label{eq:fixed_exp}
\end{align}
Hence, the linear response of $\rho(A)$ at time $T$ is given by
\begin{align*}
\delta_T \rho(A) &\approx \lim_{t \rightarrow \infty} \frac{1}{t} \sum_{k=1}^t \int l(dx) \sum_{j=0}^{k-1} X(T-j,f^{k-j}(x)) D(A \circ f^{j}) (f^{k-j}(x)) \nonumber  \\
%& = \lim_{t \rightarrow \infty} \frac{1}{t} \sum_{j = 0}^{t-1} \sum_{i=1}^{t-j}  \int l(dx) X(T-j,f^i(x)) D(A \circ f^{j}) (f^i(x)) \nonumber \\
& = \lim_{t \rightarrow \infty} \frac{1}{t} \sum_{j \geq 0} \int  \sum_{i=1}^{t-j} (f^*)^{i} l(dx) X(T-j, x) D(A \circ f^{j}) (x) \,. \nonumber \\
\end{align*}
Note that in contrast to Eq.~\ref{eq:future} the indices are such that the time average of the measure and the perturbation are decoupled. This crucially depends on the choice of the sampling. This allows us to use the definition of the SRB measure in Eq.~\ref{eq:SRB} and replace the time average in the limit by $\rho$:
\begin{align}
\delta_T \rho(A) & \approx \sum_{j \geq 0} \int  \rho(dx) X(T-j,x) D(A \circ f^{j}) (x) \,. \label{eq:lin_response}
\end{align}
This agrees with Eq.~\ref{eq:ruelle_time}. Here we do get the anticipated result. Note that this expression gives also a non-zero response for a perturbation which is non-zero only for a finite time, as opposed to Eq.~\ref{eq:future}.

This sampling is the natural one fore deducing the general linear response theory. Doing the calculation for constant forcing does not elucidate the relevance of the choice of sampling. This sampling corresponds to a Gedankenexperiment where the system is prepared in the distant past and we observe the difference of the perturbed and unperturbed evolution up to a given instant $T$.

\section{Numerics}
\label{sec:numerics}

\subsection{Reick's formula}
For perturbations that are separable ($X(t,x)=\phi(t)\chi(x)$) and
have a single driving frequency $\Omega$ ($\phi(t)=\epsilon
cos(\Omega t)$), the
following sampling scheme for computing the susceptibility for a
given observable $A$ has been proposed by Reick~\cite{reick_linear_2002}:
\begin{align}
\hat{\kappa}_A(\Omega) = \lim_{\epsilon \rightarrow 0} \lim_{N \rightarrow
\infty} \frac{1}{N \epsilon} \sum_{t=1}^{N } e^{i \Omega
t} \int \rho(dx) \left( A( \tilde{f}_0^t (x)) - A(f^t (x))
\right). \label{eq:reick}
\end{align}
This formula has been later adopted to analyze the output of a
simple climate model \cite{luc11} and a generalization has been
proposed to study the nonlinear susceptibilities describing
harmonic generation \cite{luc09}. Applicability of this formula depends on performing numerical experiments where the initial samples approximate the unperturbed SRB measure $\rho$.

Using our previous calculations, we want to circumstantiate the validity of the
formula. We apply Ruelle's response theory to obtain a perturbative expression of Eq.~\ref{eq:reick} in terms of quantities of the unperturbed dynamics.
\begin{align*}
& \lim_{N\rightarrow \infty} \frac{1}{N } \sum_{t=1}^{N } e^{i \Omega t} \int l(dx) \left( A(\tilde{f}^t_0(x)) - A(f^t(x)) \right) \nonumber \\
&\approx \lim_{N\rightarrow \infty} \frac{1}{N } \sum_{t=1}^{N } e^{i \Omega t} \int \sum_{j=1}^t (f^j)^* l(dx) X(j,f^j(x)) D( A \circ f^{t-j}) (x) \,. \nonumber \\
\end{align*}
Here we encounter the same problem as in Eq.~\ref{eq:future}, namely the coupling of the averages of the measure and the perturbation. Indeed, using Reick's formula sampling from an initial measure different from the unperturbed SRB measure, one does not get a reasonable response, as reported in~\cite{luc11}.

By sampling according to the unperturbed SRB measure $\rho$ instead of $l$, the above equation becomes
\begin{align}
&\lim_{N\rightarrow \infty} \frac{1}{N } \sum_{j=1}^{N }  \sum_{k=0}^{N  - j} e^{i \Omega k} e^{i \Omega j} \int \rho(dx) X(j,x) D( A \circ f^{k}) (x) \nonumber \\
% &= \lim_{N\rightarrow \infty} \frac{1}{N } \int \rho(dx) \sum_{k \geq 0} e^{i \Omega k} D(A \circ f^k)(x) \sum_{j=1}^{N  -k} e^{i \Omega j} X(j,x) \nonumber \label{eq:step1}
&= \lim_{N\rightarrow \infty} \frac{1}{N } \int \rho(dx) G_A(\omega,x) \sum_{j=1}^{N} e^{i \Omega j} X(j,x) \,. \label{eq:step1}
\end{align}
We insert the inverse discrete time Fourier transform
\begin{align*}
X(j,x) = \frac{1}{2 \pi} \int_{-\pi}^\pi \hat{X}(\omega,x) e^{-i \omega j} d\omega
\end{align*}
into Eq.~\ref{eq:step1}:
\begin{align}
&\int \rho(dx) G_A(\omega,x) \lim_{N \rightarrow \infty} \frac{1}{N } \sum_{j=1}^{N } e^{i \Omega j} X(j,x) \nonumber \\
&= \int \rho(dx) G_A(\omega,x) \lim_{N \rightarrow \infty} \frac{1}{N } \sum_{j=1}^{N } e^{i \Omega j} \frac{1}{2 \pi} \int_{-\pi}^\pi \hat{X}(\omega,x) e^{-i \omega j} d\omega \nonumber \\
&= \int \rho(dx) G_A(\omega,x) \lim_{N \rightarrow \infty} \frac{1}{2\pi} \int_{-\pi}^\pi \hat{X}(\omega,x) u_N(\Omega - \omega)  d\omega \nonumber \\
&= \lim_{N \rightarrow \infty} \frac{1}{2 \pi} \int_{-\pi}^\pi d\omega u_N(\Omega-\omega) \hat{\kappa}_A (\omega)
\,. \label{eq:lim1}
\end{align}
where
\[u_N(\Omega-\omega)=\frac{1}{N } \sum_{j=1}^{N} e^{i(\Omega - \omega)j} \]
This can be rewritten by making use of
\begin{align*} x + \ldots + x^N = x \frac{1-x^N}{1-x} \,. \end{align*}
as
\begin{align*}
u_N(\Omega-\omega)=\frac{1}{N} e^{i(\Omega-\omega)} \frac{1-e^{i N  (\Omega-\omega)}}{1-e^{i (\Omega-\omega)}} \,,
\end{align*}
which converges to $0$ as $N$ goes to infinity, except for $\Omega=\omega$. At $\Omega=\omega$ the sum over $j$ gives $N$. Hence, if $\hat{X}(\omega,x)$ is integrable, we can take the limit in Eq.~\ref{eq:lim1} inside the integral
\begin{align*}
\frac{1}{2\pi} \int_{-\pi}^\pi \hat{X}(\omega,x) \mathbf{1}_{\{ \Omega \} } (\omega) d\omega = 0 \,,
\end{align*}
where $\mathbf{1}_{\{ \Omega \} }$ is the indicator function on $\{ \Omega \}$. We deduce that in the case of a general perturbation with a
continuous Fourier spectrum, Reick's numerical approach cannot be
applied. Note also that for finite time steps $N$ (as is always the case for numerical experiments), there is an additional broadening of the signal of order $1/N$, as is apparent from Eq.~\ref{eq:lim1}.

If on the other hand the Fourier transform is singular, for example
\begin{align*}
X(\omega,x)&=\delta(\Omega-\omega) \chi(x) \,,
\end{align*}
which corresponds to the monochromatic signal
\begin{align*}
X(j,x) &= \frac{1}{2 \pi} e^{-i \Omega j} \chi(x) \,,
\end{align*}
we have that
\begin{align*}
\lim_{N\rightarrow \infty} \frac{1}{N} \sum_{j=1}^{N} e^ {i \Omega j} X(j,x) &= \frac{\chi(x)}{2 \pi} \,.
\end{align*}

Therefore, Eq. \ref{eq:step1} becomes
\begin{align*}
\int \rho(dx) G_A(\omega,x) \chi(x) = \hat{\kappa}_A(\Omega) \,,
\end{align*}
as predicted by Reick.

The above calculation shows how the explicit expansion of Reick's response formula allows us to interpret its finite time behaviour. Eq.~\ref{eq:lim1} shows how this sampling scheme amounts to filtering the susceptibility $\hat{\kappa}_A$ with the function $u_N$.

Note that in the case of a several frequencies contributing to the forcing, we are again in the case of general periodic forcing. The susceptibility can in this case be computed in two ways. Either one uses Reick's formula at every frequency present in the signal, which amounts to doing spectroscopy. On the other hand, one can also compute the full response $\delta_T \rho$ at all phases over on period and apply a Fourier transform to this time-dependent function. The response at any one specific phase can be efficiently computed with the periodic sampling strategy proposed in Eq.~\ref{eq:periodic_sampling}. In this approach each value  for the difference of $A$ between perturbed and unperturbed is processed only once, compared to the summation being done for every frequency with Reick's formula.

\subsection{Sampling continuous spectra}

The discussion in the previous subsection demonstrates how sampling according to Formula \ref{eq:reick} can only give a correct result in cases where the Fourier spectrum of the perturbation is discrete. In this subsection we explore how the discussion on the expansion at a fixed time horizon can help us find a sampling for the case of a continuous Fourier spectrum.

One possibility is to sample the response directly from the full formula of the perturbation of the SRB measure:
\begin{align*}
\delta_T \rho (A) = \lim_{t \rightarrow \infty} \frac{1}{t} \sum_{k=1}^t \int l(dx) A(\tilde{f}_{T} \circ \ldots \circ \tilde{f}_{T-k} (x)) - A(f^k (x))
\end{align*}
This sampling however entails some practical difficulties. As can be seen from the formula, an ergodic average is taken over the length of the numerical run $k$. Increasing $k$ to $k+1$ is equivalent to altering the initial conditions from $x$ to $\tilde{f}_{T-k-1}(x)$. This trajectory cannot be recovered from the previously calculated trajectories of length $k$. Hence one needs to redo the calculations of the trajectories for every value of $k$. As we will see, less costly sampling methods can be devised.

To study the behaviour of different sampling methods, let us define the following quantity:
\begin{align*}
\delta^{(k,n)} \rho_T (A) = \int l(dx) A( \tilde{f}_{T} \circ \ldots \circ \tilde{f}_{T-k} \circ f^n (x) ) - A(f^{k+n}(x))
\end{align*}
By changing $k$, we control the length of time over which we observe the difference between the perturbed and unperturbed dynamics. The initial measure $l$ is furthermore transformed by $n$ applications of the unperturbed dynamics. By increasing $n$, the initial measure $l$ converges to the unperturbed SRB measure $\rho$.

Making use of Equation~\ref{eq:fixed_exp}, we can expand $\delta^{(k,n)} \rho_T$ to get a better idea of the behaviour of this quantity under different limits and ergodic averages:
\begin{align}
\delta^{(k,n)} \rho_T (A) = \sum_{j=0}^{k-1} \int {f^{(k-j+n)}}^*l(dx) X(T-j,x) D(A\circ f^j)(x) \label{eq:expanded_diff}
\end{align}
The aim when constructing a sampling scheme is to take limits and ergodic averages over $k$ and $n$ in such a way that the response given by Equation~\ref{eq:ruelle_time} is obtained. As we have seen with Reick's formula, this convergence can depend on the perturbation $X$. Furthermore, as exemplified by the discussion in this section, numerical cost should be considered. The following ergodic mean:
\begin{align}
 \lim_{t \rightarrow \infty} \frac{1}{t} \sum_{k=1}^t \delta^{(k,n)} \rho_T (A) \label{eq:k_ergodic}
\end{align}
converges to the response $\delta_T \rho(A)$ given by Eq.~\ref{eq:expanded_diff} for any value of $n$. This can be shown following the discussion provided in Section \ref{sec:backward}, where we discuss the case $n=0$. Another possibility to obtain the SRB measure $\rho$ in Eq.~\ref{eq:expanded_diff} is to take the limit of $n$ going to infinity for a fixed $k$. We obtain: 
\begin{align}
\lim_{n \rightarrow \infty}\delta^{(k,n)} \rho_T(A) &= \sum_{j=0}^{k-1} \int \rho(dx) X(T-j,x) D(A\circ f^j)(x) \label{eq:expanded_diff2}.
\end{align}
where we have assumed that $\lim_{m\rightarrow \infty} {f^m}^* l = \rho$. This expression tends to $\delta_T \rho(A)$ in the limit of $k\rightarrow\infty$. From a theoretical point of view, an increase in the value of $n$ simply translates into a change in the initial measure $l$. Numerically, though, doing a long initial unperturbed run will evolve the initial measure towards the invariant measure $\rho$, hence improving convergence when Eqs. \eqref{eq:expanded_diff}-\eqref{eq:expanded_diff2} are considered. In fact, there are a number of different options when attempting to reach a good numerical convergence. These include
\begin{itemize}
\item increasing the length of unperturbed and perturbed trajectories ($n$ and $k$)
\item enlarging the number of initial conditions (chosen according to $l$)
\item deciding whether or not ergodic averaging is performed over $n$ and $k$.
\end{itemize}
In the limit of infinitely long perturbed runs, these approaches give the same result. However, for finite time they will perform differently.

%\begin{figure}
%\center
%\label{fig:timeline}
%\includegraphics[scale=1]{timeline}
%\caption{The different methods of calculating the linear response. In blue: the moving time horizon, in green: the fixed time horizon.}
%\end{figure}

\section{Conclusions}
In this paper we have reconsidered Ruelle's linear response theory by analyzing the impact of choosing different methods of sampling in relation to different classes of forcings. Explicitly doing an expansion of the perturbed dynamics around the unperturbed measures allows us to explore which sampling methods converge and under which conditions.

The general response formula is obtained by choosing a specific sampling where the system is prepared in the distant past and we observe the difference of the perturbed and the unperturbed dynamics up to a given time $T$. By proposing a general decomposition of space-time dependent forcings using a Schauder decomposition, we have elucidated that it is possible to define \emph{elementary} linear susceptibilities that allow to construct the response of the system to any pattern of forcing.

The other possible sampling strategy, where the time horizon is not fixed, does not give rise to a natural response theory except for constant perturbations. In the case of periodic forcings one can obtain a meaningful formula by redefining appropriately the response, finely tuned to the forcing under investigation. One needs to subsample the signal with the same period of the forcing and explore all the initial phases. By taking this approach, it is in principle possible to discover the fundamental period of the external perturbations by varying the sampling period. Thanks to our approach we get a deeper understanding of the range of applicability of Reick's formula, which has been used as a signal processing tool to study the linear response of numerical models.

Nonetheless, this approach fails if the forcing is not periodic, in which case we must resort to the fixed-time horizon framework to get a meaningful answer. In fact, our findings explain why considering the fixed-time horizon it is possible to analyze a response to forcings that have a continuous Fourier spectrum. The clarifications presented in this paper may be of relevance for devising the data processing for actual laboratory experiments on nonlinear systems.

We also clarify that it is crucial in practical terms to use an ensemble approach where the initial conditions sample approximately the unperturbed SRB measure. Our calculation is explicitly performed for discrete time, but the analogous results for continuous time are presented in Appendix~\ref{app:continuous}. Moreover our considerations seem to be appropriate also for the case of nonlinear response~\cite{ruelle_d._nonequilibrium_1998,luc09}.

To summarize, we have shown the following
\begin{itemize}
\item Sampling a general response from an initial time up to a moving time horizon does not lead to a well-defined sampling method.
\item Starting the simulation at times in the distant past and averaging the response at a fixed time horizon always results in the full response of the system at the fixed point in time. This approach can be computationally inefficient.
\item Sampling a periodic response with a moving time horizon results in a response of the system as if it were forced with an averaged forcing.
\item In the periodic case, the full response can be computed by sampling with a horizon moving forward in time with steps of one period. This response depends on the initial phase. The susceptibility can be computed through a Fourier transform of the response. 
    \item A constant forcing can be considered as a periodic forcing with period $1$ and can thus be sampled with a horizon moving with time steps of $1$.
\item For periodic forcings, Reick's spectroscopic formula also allows to discern the response at different frequencies, i.e. the susceptibility. It gives a zero susceptibility for forcings with a continuous spectrum.
\end{itemize}

We believe that the results presented in this article can be of interest to researchers interested in studying the response of complex systems to modulations of their internal parameters or to external perturbations. For various reasons, climate science is an especially promising field of application. First of all, we clarify crucial differences between sampling periodic and aperiodic forcings. This is a crucial issue if one wants to apply linear response theory to study different scenarios such as the response of the system to monotonically increasing CO$_2$ levels (see a forthcoming paper by the authors) versus its response to periodic forcings such as those due to astronomical and astrophysical phenomena. In particular, we have proposed a parsimonious but effective way for analysing periodic - but non-monochromatic - forcings. Again, this setting is applicable to climate science due to the presence of cycles with different time scale, such as the daily, yearly and solar cycles.

Future work will address the investigation of the response of a non-equilibrium system to a general random field. Moreover, we will analyze the impact of the various sampling schemes described in this paper when studying the output of numerical models. 

\section*{\small{Acknowledgements}}
VL, JW, DF acknowledge the financial support of the EU-ERC project NAMASTE-Thermodynamics of the climate system. TK acknowledges F. Bonetto for fruitful discussions.

\appendix
\section{Continuous time response formulas}
\label{app:continuous}
Here we give the formulas for continuous time systems corresponding to the ones presented in the main text. The time evolution is in this setting given by a differential equation
\begin{align*}
\frac{dx}{dt} = F(x) \,,
\end{align*}
resulting in a flow $ x(t+s) = f^s (x(t)) $.
The SRB measure is given by
\begin{align*}
\rho(A) = \lim_{t \rightarrow \infty} \frac{1}{t} \int_0^t ds \int l(dx) A(f^s(x)) \,.
\end{align*}
For a separable perturbation
\begin{align*}
\frac{dx}{dt} = F(x) + \chi(x) \phi(t)
\end{align*}
the susceptibility is given by
\begin{align*}
\hat{\kappa}_A(\omega) = \int_0^\infty dt e^{i \omega t} \int \rho(dx) \chi(x) D(A \circ f^t)(x) \,.
\end{align*}
In case of a general perturbation $F(x) \rightarrow F(x) + X(t,x)$ the linear response becomes:
\begin{align*}
\delta_T \rho(A) \approx \int_0^\infty d\tau \int \rho(dx) X(T-\tau,x) D(A \circ f^\tau)(x)
\end{align*}

The SRB measure with a moving time horizon is:
\begin{align*}
\tilde{\rho}_T = \lim_{t \rightarrow \infty} \frac{1}{t} \int_0^t ds \int l(dx) A( \tilde{f}^{T+t}_T (x))
\end{align*}
and the SRB measure with a fixed time horizon:
\begin{align*}
\tilde{\rho}_T = \lim_{t \rightarrow \infty} \frac{1}{t} \int_0^t ds \int l(dx) A( \tilde{f}^{T}_{T-t} (x))
\end{align*}
where $\tilde{f}_{t_1}^{t_2}(x)$ is a trajectory of the perturbed system, starting at time $t_1$ in $x$ and evolving up to time $t_2$.

Reick's formula now becomes:
\begin{align*}
\hat{\kappa}_A(\omega) = \lim_{\epsilon \rightarrow 0} \lim_{\nu \rightarrow \infty} \frac{1}{\nu \epsilon} \int_0^\nu dt e^{i \Omega t} \int \rho(dx) \left( A(\tilde{f}_0^t(x)) - A(f_0^t(x)) \right)
\end{align*}

\newpage
\bibliographystyle{plain}
\bibliography{response}
\newpage
\begin{figure}
\center
\includegraphics[scale=1]{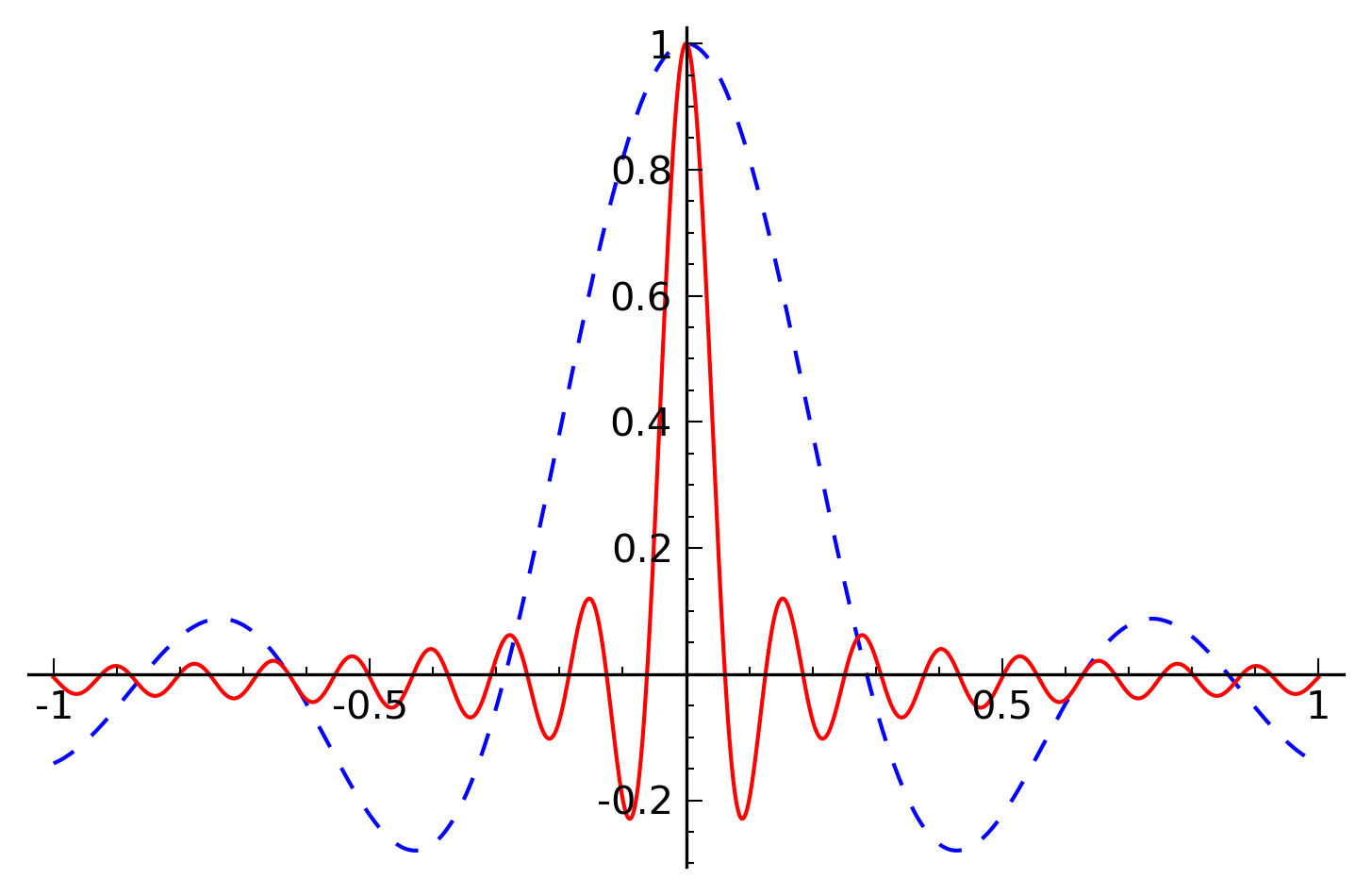}
\caption{The convergence of $u_N$ to the indicator function $\mathbf{1}_{ \{ 0 \} }$. The dashed blue line shows $u_{10}$, the full red line $u_{50}$.}\label{fig:u_N}
\end{figure}
\end{document}